# Describe NMR relaxation by effective phase diffusion equation

Guoxing Lin*

*Carlson School of Chemistry and Biochemistry, Clark University, Worcester, MA 01610, USA*

*Email: glin@clarku.edu

**Abstract**

This paper proposes an effective diffusion equation method to analyze nuclear magnetic resonance (NMR) relaxation. NMR relaxation is a spin system recovery process, where the evolution of the spin system is affected by the random field due to Hamiltonians, such as dipolar couplings. The evolution of magnetization can be treated as a random walk in phase space described either by a normal or fractional phase diffusion equation. Based on these phase diffusion equations, the NMR relaxation rates and equations can be obtained, exemplified in the analysis of relaxations affected by an arbitrary random field, and by dipolar coupling for both like and unlike spins. The obtained theoretical results are consistent with the reported results in the literature. Additionally, the anomalous relaxation expression obtained from the Mittag-Leffler function based time correlation function can successfully fit the previously reported $^{13}$C $T_1$ NMR experimental data of polyisobutylene (PIB) in the blend of PIB and head-to-head poly(propylene) (hhPP). Furthermore, the proposed phase diffusion approach provides an intuitive way to interpret NMR relaxation, particularly for the fractional NMR relaxation, which is still a challenge to explain by the available theoretical methods. The paper provides additional insights into NMR and magnetic resonance imaging (MRI) relaxation experiments.

1. **Introduction**

Nuclear magnetic resonance (NMR) relaxation is a powerful technique for detecting molecular dynamics [1,2,3] in many systems such as biological or polymer systems [4,5,6]. NMR relaxation is a recovery process where the spin system's population returns to equilibrium after being perturbed. The relaxation rate depends on the molecular motion, which changes relative molecular orientations and thus modulates many fundamental Hamiltonians, including dipolar coupling, quadrupolar coupling, chemical shift anisotropy, etc. [1] These modulated Hamiltonians can be viewed as random fluctuating fields alerting the spin system. The perturbation affects the spin evolution and changes the spin relaxation process, which is often treated by the density operator theory based on quantum mechanics. Traditional theories such as Bloch-Wangsness-Redfield theory and the second-order perturbation theory have successfully explained many normal NMR relaxation processes [1,2].

However, NMR relaxation in polymer and biological systems is often anomalous. Anomalous NMR relaxation could arise from various aspects. First, the relative molecular motion modulating the random field can be anomalous. The monoexponential correlation function based on normal rotational or translational diffusion is insufficient to describe the anomalous relative motion. Thus, fractional rotational or translational diffusion has been employed to describe the anomalous relative motion for NMR relaxation and other relaxation processes [7, 8,9,10,11,12]. The time correlation function of the random fluctuating field from these reports is often a Mittag-Leffler function $E_{\alpha,1}\left(-\left(\frac{t}{\tau}\right)^\alpha\right)$, or a stretched exponential function $\exp\left(-\left(\frac{t}{\tau}\right)^\alpha\right)$, where $\alpha$ is the order of time-fractional derivative, and $\tau$ is the characteristic time. The Mittag-Leffler function can be reduced to a stretched exponential function when $\left(\frac{t}{\tau}\right)^\alpha$ is small. The stretched exponential function $\exp\left(-\left(\frac{t}{\tau}\right)^\alpha\right)$ is the same as the Kohlrausch-Williams-Watts (KWW) function [13,14], a commonly used time correlation function for macromolecular systems. Second, the relaxation itself(evolution of magnetization of the investigated spin system) is anomalous as described by fractional Bloch-equation proposed by Ref. [15]. The fractional Bloch-equation yields a Mittag-Leffler function-based NMR relaxation. However, interpreting such a fractional NMR relaxation behavior is challenging for traditional theoretical methods, such as the perturbation theory. Therefore, it is necessary to develop a theoretical treatment to describe fractional NMR relaxation behavior.

To better understand the normal and anomalous NMR relaxations, a phase diffusion approach is proposed to describe NMR relaxation. The NMR relaxation process is the evolution of spin systems affected by random Hamiltonians, as mentioned previously. It can be described either by the evolution of wavefunctions (Schrödinger representation) or the evolution of observable operators (Heisenberg representation) such as angular momentum [1,2]. Under the random field's influence, the phase of the angular momentum undergoes a random motion, which could be seen as a random walk in phase space and consequently be described by a phase diffusion equation. The phase diffusion equation proposed here is similar to the effective phase diffusion equation recently developed to describe the phase evolution of spin coherence affected by pulsed-field gradient (PFG) [16]. By solving the diffusion equation, the phase distribution of the magnetization can be obtained, and the relaxation of magnetization can be determined.

The phase diffusion describing the magnetization evolution can either be a normal diffusion or an anomalous diffusion. Both normal and anomalous phase diffusion equations describing NMR relaxation will be investigated in this paper. The anomalous diffusion can be modeled by fractional calculus [18, 17, 18, 19,20]. Based on continuous-time random walk (CTRW) simulation, the fractional diffusion arises from its diffusion waiting time behaving asymptotically to a power-law or its diffusion jump length following a power-law [21,22]. Normal diffusion can be viewed as a specific case of anomalous diffusion. Fractional diffusion equation has been applied to describe various anomalous dynamics processes such as anomalous NMR relaxation and PFG anomalous diffusion [8,12,16,23,24,25,26,27,28]. The time-dependent behaviors of normal dynamics processes are often described by monoexponential function. However, Mittag-Leffler or stretched exponential based function is frequently applied to describe anomalous dynamics processes such as those found in PFG signal attenuation, the time correlation function, and the NMR relaxation in polymers. . Based on the proposed phase diffusion equation, the relaxation rates and relaxation equation due to various Hamiltonians can be obtained. The obtained normal relaxation results agree with the reported theoretical result in literature [1,2], and the anomalous relaxation expression of dipolar coupling between unlike spins can successfully fit the reported $^{13}C$ $T_1$ NMR experimental data taken from Ref. [6]; additionally, the relaxation results from the anomalous phase diffusion agree with Ref. [15].

The rest of the paper is organized as follows. First, Section 2.1 gives a fundamental derivation of the phase diffusion equation that focuses on the normal phase diffusion: The relaxation effects of the simple random field and dipolar coupling are illustrated by normal phase diffusion in Section 2.1.1 and 2.1.2, respectively, and the normal phase diffusion whose jump time is determined by the fractional relative motion discussed in section 2.1.3. Secondly, in Section 2.2, the anomalous NMR relaxation arising from fractional phase diffusion is studied. The proposed approach provides an intuitive physical explanation of the NMR relaxation process. The results here give additional insights into the anomalous NMR relaxation, which could improve the analysis of NMR and magnetic resonance imaging (MRI) experiments in polymer and biological systems.

## 2. Theory

The spin system's relaxation is affected by the Hamiltonians, which are modified by the random relative motion of particles in the spin system. The relative random motion can be theoretically modeled by normal or fractional diffusion: spherical rotational diffusion (Debye diffusion) or translational diffusion [1,8,12]. While, the evolution of spin magnetization is significantly different from the relative random motion because its longitudinal component is constantly subjected to the thermal relaxation, and the spin moment is constantly rotating around the external magnetic field. The Hamiltonian interaction effect modified by the spin system's relative motion can be viewed as a random fluctuating field. The fluctuating field makes the spin magnetization undergo random phase motion that could be treated by a diffusion equation. Regardless of a normal or fractional relative random motion of spins, a finite effective average jump waiting time $\tau_{jump}(\omega)$ (existing a finite first moment of time distribution) in the rotating frame leads to a normal phase diffusion; in contrast, in a complex system, a power-law distribution of $\tau_{jump}(\omega)$ could lead to a fractional phase diffusion. The normal phase diffusion will be studied in section 2.1, while the fractional phase diffusion will be investigated in section 2.2.

2.1 NMR relaxation described by normal phase diffusion

2.1.1 Random field

The random molecular motion alerts the relative molecular orientations, modulating many fundamental Hamiltonians of spin systems. These interactions can be viewed as a random field exerting on the observed spin moments. A simplified random field $H_1(t)$ can be used as a basic model to demonstrate how a random field affects the spin relaxation. $H_1(t)$ can be given by [2]

$$H_1(t) = \sum_{q=x,y,z} H_q G(t) I_q, \tag{1}$$

where $I_q$ is the component of the angular momentum, and $G(t)$ is time correlation function resulting from the relative rotational or translational motion of the lattice particles, and $H_q$ is the amplitude of the random field described by

$$H_q = \gamma \hbar h_q, \tag{2}$$

where $\gamma$ is the gyromagnetic ratio, $\hbar$ is the reduced Planck constant, and $h_q$ is the magnetic field intensity.

First, let us see the effect the component $H_x(t)I_x$, which causes the angular momentum $I_z$ to rotate according to [1,2,3]

$$I_z \xrightarrow{e^{iI_x\phi}I_z e^{-iI_x\phi}} I_z \cos\phi + I_y \sin\phi, \tag{3}$$

where

$$\phi = \int_0^t \frac{H_x(\tau)}{\hbar} d\tau = \int_0^t \gamma h_x G(\tau) d\tau. \tag{4}$$

Eqs. (3) and (4) indicate that $I_z$ undergoes a rotation $e^{-i\phi}$ in $I_z$ and $I_y$ plane. For convenience, such a random rotation can be represented in $I_z$ and $I_y$ plane as

$$I_z \xrightarrow{e^{iI_x\phi}I_z e^{-iI_x\phi}} I_z e^{-i\phi}, \tag{5}$$

and the rotating phase $\phi$ in Eq. (4) can be rewritten in a discrete mode as

$$\Delta\phi_i = \gamma h_x G(t_i) \Delta t_i, \tag{6a}$$

$$\phi = \sum_{i=1}^n \gamma h_x G(t_i) \Delta t_i, \tag{6b}$$

where $\Delta\phi_i$ is the phase jump length, $\Delta t_i$ is the jump time, and $t_i = \sum_{j=1}^i \Delta t_j$. Eqs. (6a) and (6b) indicate a random phase walk, a phase diffusion process.

The time correlation function $G(t)$ arising from the random relative motion relates the Hamiltonians at different instants, here at time zero and at time *t*. A typical time correlation function is the monoexponential function [4,5,13,14]

$$G(t) = \exp\left(-\frac{t}{\tau}\right), \tag{7}$$

where $\tau$ is the reorientation time constant. In a polymer or macromolecular system, such a monoexponential correlation time corresponds to a mean rotational motion time:

$$\int_0^\infty G(t)dt = \int_0^\infty \exp\left(-\frac{t}{\tau}\right)dt = \tau. \tag{8}$$

Because the correlations existing before and after time zero are equivalent, $2\tau$ could be viewed as an average jump time $\tau_{jump0}$ of the random phase walk, namely

$$\tau_{jump0} = \int_{-\infty}^\infty G(t)dt = \int_{-\infty}^\infty \exp\left(-\frac{|t|}{\tau}\right)dt = 2\tau, \tag{9}$$

The average phase jump length is

$$\langle \Delta\phi_i \rangle = \omega_x \tau_{jump0} = \omega_x 2\tau, \tag{10}$$

where $\omega_x$ is the angular frequency described by

$$\omega_x = \frac{H_q}{\hbar} = \frac{\gamma \hbar h_x}{\hbar} = \gamma h_x. \tag{11}$$

The phase diffusion constant can be obtained by [16]

$$D_{\phi x} = \frac{\langle (\Delta\phi)^2 \rangle}{2\tau_{jump0}} = \frac{\omega_x^2 \tau_{jump0}}{2} = \frac{\omega_x^2}{2} \int_{-\infty}^\infty G(t)dt \tag{12}$$

$D_{\phi x}$ will be referred to as static diffusion constant, $D_{\phi x,static}$, because it does not consider that the investigated spin system is continuously rotating around the external field $H_0$ at Larmor frequency $\omega_0$. For a quantum coherence of order $n$, this rotating angular frequency is

$$n\omega_0 = n\gamma H_0. \tag{13}$$

The rotation affects the observable quantities of phase diffusion. The effective diffusion coefficient can be given by

$$D_{\phi x,eff} = \varepsilon D_{\phi x,static}, \tag{14}$$

where $D_{\phi x,static}$ is given by Eq. (12) as mention above, and $\varepsilon$ is the effective coefficient described by

$$\varepsilon = \frac{\int_{-\infty}^\infty G(t)\exp(-in\omega_0 t)dt}{\int_{-\infty}^\infty G(t)dt}. \tag{15}$$

The substitution of $D_{\phi x,static} = D_{\phi x}$ and $\varepsilon$ into Eq. (14) gives

$$D_{\phi x,eff} = \varepsilon D_{\phi x,static} = \frac{\omega_x^2}{2} \int_{-\infty}^\infty G(t)\exp(-in\omega_0 t)dt = \omega_x^2 \tau \frac{1}{1+(n\omega_0\tau)^2}. \tag{16}$$

When $\omega_0 = 0$, $D_{\phi x,eff}$ reduces to $D_{\phi x}$ (namely $D_{\phi x,static}$) in Eq. (12).

Another way to understand the change of diffusion affected by the rotation is to introduce an effective average jump time of the random phase walk, $\tau_{jump}(\omega)$, which can be given by

$$\tau_{jump}(\omega) = \int_{-\infty}^\infty G(t)\exp(-i\omega t)dt; \tag{17a}$$

when $G(t) = \exp\left(-\frac{t}{\tau}\right)$,

$$\tau_{jump}(n\omega_0) = \frac{2\tau}{1+(n\omega_0\tau)^2}. \tag{17b}$$

The effective diffusion constant in the rotating frame can be given by

$$D_{\phi x,eff} = \frac{\langle(\Delta\phi)^2\rangle\varepsilon}{2\,\tau_{jump0}} = \frac{\omega_x^2 \tau_{jump0}\varepsilon}{2} = \frac{\omega_x^2 \tau_{jump}(n\omega_0)}{2}, \quad (18)$$

which gives the same diffusion constant as Eq. (16).

With $D_{\phi x,eff}$, the normal phase diffusion equation can be described by [16]

$$\frac{dP_z(\phi,t)}{dt} = D_{\phi x,eff}\Delta P_z(\phi,t). \quad (19)$$

Applying Fourier transform $p_z(k,t) = \int_{-\infty}^{\infty} P_z(\phi,t)\exp(-ik\phi)d\phi$ to both sides of Eq. (19), we get

$$\frac{dp_z(k,t)}{dt} = -k^2 D_{\phi x,eff} P_z(k,t). \quad (20)$$

The solution of Eq. (20) is

$$p_z(k,t) = A\exp(-D_{\phi x,eff} k^2 t), \quad (21)$$

where $A$ represents the magnetization $\langle I_z(0)\rangle$ at time zero. When $k = 1$,

$$p_z(1,t) = A\exp(-D_{\phi x,eff} t). \quad (22)$$

$p_z(1,t)$ is the average magnetization $\langle I_z(t)\rangle$ across all possible phases, namely

$$\langle I_z(t)\rangle = p_z(1,t) = \int_{-\infty}^{\infty} P_z(\phi,t)\exp(-i\phi)\,d\phi$$

$$= A\exp(-D_{\phi x,eff} t). \quad (23)$$

Based on Eq. (23), the relaxation rate $W_x$ due to the perturbation of the random field is

$$W_x = D_{\phi x,eff} = \frac{\omega_x^2 \tau_{jump}(n\omega_0)}{2} = (\gamma h_x)^2 \frac{\tau}{1+(n\omega_0\tau)^2}. \quad (24)$$

Because of the coupling between the spin system and the lattice, a thermal relaxation process always occurs, which leads to the equilibrium population distribution [1,2]. The thermal relaxation (see **Appendix A**) can be included in Eq. (20); when $k = 1$, Eq. (20) can be modified as

$$\frac{dp_z(1,t)}{dt} = -D_{\phi x,eff}\bigl(p_z(1,t) - p_{z0}(1,0)\bigr), \quad (25)$$

where $p_{z0}(1,0)$ is the probability distribution function representing the equilibrium magnetization $M_{z0}$. Eq. (25) can be rewritten as

$$\frac{d\langle M_z\rangle}{dt} = -\frac{1}{T_1}(\langle M_z\rangle - M_{z0}), \quad (26)$$

where $M_z$ is the magnetization, and the spin-lattice relaxation time $T_1$ is described as

$$\frac{1}{T_1} = D_{\phi x,eff}. \quad (27)$$

Eq. (26) is the familiar spin-lattice relaxation equation [1,2,3].

There are still two other components of the random field, $H_y(t)$ and $H_z(t)$ in Eq. (1) need to be considered. $H_z(t)$ commutes with $I_z$, so it does not affect the evolution of $I_z$ [1,2]. While the effect of the component $H_y(t)$ on the angular momentum $I_z$ is similar to that of $H_x(t)$, and thus its corresponding phase diffusion constant can be obtained similarly as $D_{\phi y,eff} = \frac{\omega_y^2 \tau_{jump}(n\omega_0)}{2} = (\gamma h_y)^2 \frac{\tau}{1+(n\omega_0\tau)^2}$. Because $H_x(t)$ and $H_y(t)$ are two uncoupled random interactions, their combined effects on the NMR relaxation is obtained straightwardly as

$$\frac{1}{T_1} = D_{\phi x,eff} + D_{\phi y,eff} = \gamma^2(h_x^2 + h_y^2)\frac{\tau}{1+(n\omega_0\tau)^2}. \quad (28)$$

When $n = 1$, the single quantum coherence case, Eq. (28) exactly replicates the Redfield theory's result in Ref. [2].

For the transverse magnetizations $M_x$ and $M_y$ affected by the random field $H_1(t)$ described by Eq. (1), the observable angular momentum $I_x$ corresponding to the magnetization $M_x$ is affected by $H_z(t)$ and $H_y(t)$, while $I_y$ is affected by $H_z(t)$ and $H_x(t)$ [1,2]. Only the angular momentum $I_x$ will be considered. For simplicity, the rotating frame reference is used. Similar to the derivation of Eq. (25) for angular momentum $I_z$ affected by the random field $H_x(t)$, it is straightforward to obtain

$$\frac{dP_x(1,t)}{dt} = -D_{\phi z, eff} P_x(1,t). \tag{29a}$$

$$\frac{dP_x(1,t)}{dt} = -D_{\phi y, eff} P_x(1,t). \tag{29b}$$

In Eqs. (29), the relaxation of transverse magnetization is not affected by the lattice relaxation; therefore, no thermal relaxation correction like Eq. (25) is needed. Additionally, the average transverse magnetization $\langle M_x \rangle$ always relaxes toward zero. The combined effect of both $z$ and $y$ direction random fields upon spin-spin relaxation of the magnetization $M_x$ is

$$\frac{1}{T_{2x}} = D_{\phi z, eff} + D_{\phi y, eff} = \gamma^2 h_z^2 \tau + \gamma^2 h_y^2 \frac{\tau}{1+(n\omega_0 \tau)^2}, \tag{30a}$$

and

$$\frac{d\langle M_x \rangle}{dt} = -\frac{1}{T_{2x}} \langle M_x \rangle, \tag{30b}$$

which agrees with the results in Ref. [2].

It is worth to note that in Eq. (30a), $D_{\phi z, lab} = \gamma^2 h_z^2 \tau$, because the Hamiltonian $H_z(t)$ belongs to zero quantum coherence.

2.1.2 Dipolar coupling

The random field of a real spin system is much more complex than the simple model mentioned above. The random filed arising from dipolar coupling in the system can be written as [1,3]

$$H_d(\Omega) = \sum_q A^{(q)} F^{(q)}(\Omega), \tag{31a}$$

where $F^{(q)}(\Omega)$ are the lattice operators, $F^{(q)}(\Omega) = F^{(-q)*}(\Omega)$, and $A^{(q)}$ are the spin operators, $A^{(q)} = A^{(-q)+}$. $F^{(0)}(\Omega)$, $F^{(1)}(\Omega)$, and $F^{(2)}(\Omega)$ can be expressed based on normalized spherical harmonics as [1,3]

$$F^{(0)}(\Omega) = -\frac{1}{r^3}\sqrt{\frac{16\pi}{5}} Y_2^{(0)}(\Omega), \ F^{(1)}(\Omega) = \frac{1}{r^3}\sqrt{\frac{8\pi}{15}} Y_2^{(1)}(\Omega), \ F^{(2)}(\Omega) = \frac{1}{r^3}\sqrt{\frac{32\pi}{15}} Y_2^{(2)}(\Omega), \tag{31b}$$

and the spin operators $A^{(q)}$ is

$$\begin{aligned} A^{(0)} &= -\frac{3}{2}\frac{\mu_0}{4\pi}\gamma_I \gamma_S \hbar^2 \left\{ -\frac{2}{3} I_z S_z + \frac{1}{6}(I_+ S_- + I_- S_+) \right\}, \\ A^{(1)} &= -\frac{3}{2}\frac{\mu_0}{4\pi}\gamma_I \gamma_S \hbar^2 \{I_+ S_z + I_z S_+\}, \\ A^{(2)} &= -\frac{3}{4}\frac{\mu_0}{4\pi}\gamma_I \gamma_S \hbar^2 I_+ S_+. \end{aligned} \tag{31c}$$

The dipolar coupling effect on spin-lattice relaxation will be derived for the following two categories:

i. Like spins:

The phase diffusion due to $H_d^{(q)}(\Omega)$ can be interpreted from its $D_{\phi, eff}$. Based on Eq. (16). $D_{\phi, eff} = \omega^2 \tau \frac{1}{1+(n\omega_0 \tau)^2}$. One of the dominant factors determining $D_{\phi, eff}$ is the square of angular frequency, $\omega^2$. According to Eq. (11), $\omega_{(q)} = \frac{H_q}{\hbar}$, which indicates

$$\omega_{(q)}^2 = \frac{|H_q|^2}{\hbar^2}, \tag{32}$$

where $H_q$ is the Hamiltonian term $H_d^{(q)}(\Omega)$ of dipolar coupling. In the dipolar coupling operators Eq. (31c), only the effects of $A^{(\pm 2)}$ and $A^{(\pm 1)}$ associated with $H_d^{(2)}(\Omega)$ and $H_d^{(1)}(\Omega)$ need to be considered, because $A^{(0)}$ has no net effect on the change of the populations of like spins in different energy levels [1,2]. The processes to calculate the effects of $H_d^{(2)}(\Omega)$ and $H_d^{(1)}(\Omega)$ are similar, only $H_d^{(2)}(\Omega)$ will be illustrated in detail here.

The average effect of dipolar coupling Hamiltonian $H_d^{(2)}(\Omega)$ upon the phase diffusion depends on the correlation function $P(\Omega, \Omega_0, t)$, which represents the probability of two spins' orientation changing from orientation $\Omega_0$ at time zero to $\Omega$ at time $t$ because of the random motion. $P(\Omega, \Omega_0, t)$ can be obtained from Debye rotational diffusion as [1]

$$P(\Omega, \Omega_0, t) = \frac{1}{4\pi} \sum_{l,m} Y_l^{m*}(\Omega_0) Y_l^m(\Omega) \exp\left(-\frac{t}{t_l}\right), \tag{33}$$

where $t_l$ is the rotational time constant. $P(\Omega, \Omega_0, t)$ includes the spatial correlation function $\sum_{l,m} Y_l^{m*}(\Omega_0) Y_l^m(\Omega)$, the time correlation function $\exp\left(-\frac{t}{t_l}\right)$, and the normalization factor $\frac{1}{4\pi}$. From the spatial correlation function $\sum_{l,m} Y_l^{m*}(\Omega_0) Y_l^m(\Omega)$, the average of spatial operator $F^{(2)}(\Omega)$ of the Hamiltonian $H_d^{(2)}(\Omega)$ can be calculated as

$$\langle F^{(2)}(\Omega) \rangle = \langle F^{(2)}(\Omega) \sum_{l,m} Y_l^{m*}(\Omega_0) Y_l^m(\Omega) \rangle$$
$$= Y_2^{(-2)*}(\Omega_0) \int F^{(2)}(\Omega) Y_2^{(-2)}(\Omega) \, d\Omega$$
$$= \frac{1}{r^3} \sqrt{\frac{32\pi}{15}} Y_2^{(-2)*}(\Omega_0), \tag{34}$$

and the average of $\left|\langle F^{(2)}(\Omega) \rangle\right|^2$ can be calculated as

$$\langle \left|\langle F^{(2)}(\Omega) \rangle\right|^2 \rangle = \frac{1}{4\pi} \int \left|\langle F^{(2)}(\Omega) \rangle\right|^2 d\Omega_0 = \frac{8}{15 r^6}. \tag{35}$$

Note that $\frac{1}{4\pi}$ comes from the initial probability intensity $P(\Omega_0, 0)$, which is a normalization factor pertinent to the diffusion of spin system ensemble but not directly related to individual Hamiltonians; therefore, it is $\frac{1}{4\pi}$ rather than $\left(\frac{1}{4\pi}\right)^2$ appearing in $\langle \left|\langle F^{(2)}(\Omega) \rangle\right|^2 \rangle$.

The contribution from $-\frac{3}{4} \frac{\mu_0}{4\pi} \gamma_I \gamma_S \hbar^2 I_+ S_+$ on angular momentum operator $I_z$ is $8 \left[\frac{3}{4} \left(\frac{\mu_0}{4\pi} \gamma_I \gamma_S\right)^2 \hbar^4 \frac{I(I+1)}{3}\right]^2$, where the factor $8 = 2 \times 2 \times 2$. The first two reflects the two independent operators from $I_+$, $I_x$ and $I_y$. The second two is for two independent contributions from $S_x$ and $S_y$, and the third two is because each of the term $I_+ S_+$ affects two like spins simultaneously. The factor $\frac{I(I+1)}{3}$ comes from $\langle I_q^2 \rangle = \frac{I(I+1)}{3}$, where $I$ is the spin number. Additionally, the conjugate term $I_- S_-$ of $I_+ S_+$ produces the same effect on the relaxation as that of $I_+ S_+$, which contributes to an additional factor two. Based on the above analysis, the square angular frequency $\omega_{(2)}^2$ from the dipolar coupling can be given by

$$\omega_{(2)}^2 = \frac{|H_2|^2}{\hbar^2} = \frac{8}{15 r^6} \times 8 \times \frac{9}{16} \left(\frac{\mu_0}{4\pi} \gamma_I \gamma_S\right)^2 \hbar^2 \times \frac{I(I+1)}{3} \times 2 = \frac{8}{5} \left(\frac{\mu_0}{4\pi} \gamma_I \gamma_S\right)^2 \hbar^2 \frac{I(I+1)}{3}. \tag{36}$$

According to Eq. (16), the effective diffusion constant affected by $A^{(\pm 2)}$ is

$$D_{\phi,eff}^{(2)} = \frac{\omega_{(2)}^2}{2} \tau_{jump}(2\omega_0) = \frac{4}{5 r^6} \left(\frac{\mu_0}{4\pi}\right)^2 \gamma^4 \hbar^2 I(I+1) \tau_{jump}(2\omega_0), \tag{37}$$

where $\gamma = \gamma_I = \gamma_S$ for the like spins.

Similarly, for $A^{(\pm 1)}$, we can get

$$D_{\phi,eff}^{(1)} = \frac{1}{5 r^6} \left(\frac{\mu_0}{4\pi}\right)^2 \gamma^4 \hbar^2 I(I+1) \tau_{jump}(\omega_0). \tag{38}$$

Based on Eqs. (27), (37), and (38), we get

$$\frac{1}{T_1} = D_{\phi,eff}^{(1)} + D_{\phi,eff}^{(2)} = \frac{1}{5 r^6} \left(\frac{\mu_0}{4\pi}\right)^2 \gamma^4 \hbar^2 I(I+1) \left[\tau_{jump}(\omega_0) + 4 \tau_{jump}(2\omega_0)\right]$$

$$\xrightarrow{G(t) = \exp\left(-\frac{t}{t_2}\right)} \frac{2}{5 r^6} \left(\frac{\mu_0}{4\pi}\right)^2 \gamma^4 \hbar^2 I(I+1) \left[\frac{t_2}{1+(\omega_0 t_2)^2} + \frac{4 t_2}{1+(2\omega_0 t_2)^2}\right], \tag{39}$$

where $t_2$ replaces $t_l$ because only the second-order term of $P(\Omega, \Omega_0, t)$ is left in the spatial average carried out in Eq. (34). Eq. (39) agrees with Ref. [1].

ii. Unlike spins

For two unlike spins, the situation is more complicated. The coupling system spin populations on different energy levels could be affected by both spins [29]. Additionally, the term $I_+S_- + I_-S_+$ in $A^{(0)}$ cannot be neglected because of two different spins, and thus all three $H_d^{(q)}(\Omega), q = 0,1,2$ need to be considered. The calculation process is generally similar to that of the like spins, with a slight. From $H_d^{(2)}(\Omega)$, one has

$$\omega_{(2)}^2 = \frac{|H_2|^2}{\hbar^2} = \frac{8}{15r^6} \times 4 \times \frac{9}{16}(\frac{\mu_0}{4\pi}\gamma_I\gamma_S)^2\hbar^2 \times \frac{S(S+1)}{3} \times 2 = \frac{4}{5}(\gamma_I\gamma_S)^2\hbar^2\frac{S(S+1)}{3}, \tag{40}$$

whose factor $\frac{4}{5}$ is half of $\frac{8}{5}$ in like spins' Eq. (36) because each spin is considered individually, and

$$D_{\phi,eff}^{(2)} = \frac{\omega_{(2)}^2}{2}\tau_{jump}(\omega_I + \omega_S) = \frac{2}{5r^6}\gamma^4\hbar^2 S(S+1)\tau_{jump}(\omega_I + \omega_S). \tag{41a}$$

Similarly,

$$D_{\phi,eff}^{(1)} = \frac{1}{5r^6}(\frac{\mu_0}{4\pi}\gamma_I\gamma_S)^2\hbar^2 S(S+1)\tau_{jump}(\omega_I). \tag{41b}$$

$$D_{\phi,eff}^{(0)} = \frac{1}{15r^6}(\frac{\mu_0}{4\pi}\gamma_I\gamma_S)^2\hbar^2 S(S+1)\tau_{jump}(\omega_I - \omega_S). \tag{41c}$$

In Eq. (41b), the $D_{\phi,eff}^{(1)}$ is different from Eq. (38) only in the factor $S(S+1)$ and the angular frequency $\omega_I$. Based on Eqs. (25), (26) and (41), we have

$$\frac{dp_{Iz}(1,t)}{dt} = -D_{\phi,eff,II}(p_{Iz}(1,t) - p_{Iz0}(1,0)) - D_{\phi,eff,IS}(p_{Sz}(1,t) - p_{Sz0}(1,0)),$$

$$\frac{dp_{Sz}(1,t)}{dt} = -D_{\phi,eff,SS}(p_{Sz}(1,t) - p_{Sz0}(1,0)) - D_{\phi,eff,SI}(p_{Iz}(1,t) - p_{Iz0}(1,0)), \tag{42}$$

Where

$$D_{\phi,eff,II} = \frac{1}{T_1^{II}} = D_{\phi,eff}^{(0)} + D_{\phi,eff}^{(1)} + D_{\phi,eff}^{(2)}$$

$$= \frac{1}{15r^6}(\frac{\mu_0}{4\pi}\gamma_I\gamma_S)^2\hbar^2 S(S+1)\{\tau_{jump}(\omega_I - \omega_S) + 3\tau_{jump}(\omega_I) + 6\tau_{jump}(\omega_I + \omega_S)\},$$

$$D_{\phi,eff,IS} = \frac{1}{T_1^{IS}} = \frac{1}{15r^6}(\frac{\mu_0}{4\pi}\gamma_I\gamma_S)^2\hbar^2 I(I+1)\{-\tau_{jump}(\omega_I - \omega_S) + 6\tau_{jump}(\omega_I + \omega_S)\}, \tag{43}$$

and $D_{\phi,eff,SS}$ and $D_{\phi,eff,SI}$ obey similar expressions just by interchanging the letters $I$ and $S$. Note that $D_{\phi,eff,IS}(p_{Sz}(1,t) - p_{Sz0}(1,0))$ and $D_{\phi,eff,SI}(p_{Iz}(1,t) - p_{Iz0}(1,0))$ appear in Eq. (42), because the populations of the related coherences ($n = 0$ and $n = 2$) are affected by the relaxation from both spins [29]. The results of unlike spins are the same as traditional results [1,2,3]. The equations of unlike spin can reduce to like spin equations when $I$ equals $S$. Practical experiments often tune the radio frequency pulse to only one of the two unlike spins such as $I$ spin, and thus for another off-resonance spin $S$, $p_{Sz}(1,t) \approx p_{Sz0}(1,0)$, namely $\frac{dp_{Sz}(1,t)}{dt} = 0$. Therefore, we get

$$\frac{dp_{Iz}(1,t)}{dt} = -D_{\phi,eff,II}(p_{Iz}(1,t) - p_{Iz0}(1,0)), \tag{44a}$$

and the corresponding $T_1$

$$\frac{1}{T_1} = D_{\phi x,lab,II} = \frac{1}{15r^6}(\frac{\mu_0}{4\pi}\gamma_I\gamma_S)^2\hbar^2 S(S+1)\{\tau_{jump}(\omega_I - \omega_S) + 3\tau_{jump}(\omega_I) + 6\tau_{jump}(\omega_I + \omega_S)\}$$

$$= \frac{2}{15r^6}(\frac{\mu_0}{4\pi}\gamma_I\gamma_S)^2\hbar^2 S(S+1)\left\{\frac{t_2}{1+[(\omega_I-\omega_S)t_2]^2} + \frac{3t_2}{1+(\omega_I t_2)^2} + \frac{6t_2}{1+[(\omega_I+\omega_S)t_2]^2}\right\}. \tag{44b}$$

Based on Eq. (44b), for $^{13}C$ spin lattice relaxation experiment,

$$\frac{1}{T_1} = n_H D_{\phi x, lab, II}, \tag{44c}$$

where $n_H$ is the number of attached Hydrogen. Monoexponential $G(t) = \exp\left(-\frac{t}{t_2}\right)$ is used to evaluate $\tau_{jump}(\omega)$ in all the three terms inside the curly bracket. A quite different, anomalous $G(t)$ based on Mittag-Leffler function will be discussed in the subsequent section.

2.1.3 Normal phase diffusion with fractional relative motion

In sections 2.1.1 and 2.1.2, the time correlation function employed is monoexponential because the relative motion of spin and lattice particles is described by normal rotational or translational diffusion in real space. However, the monoexponential function is often not sufficient to explain relaxation experiments in macromolecular systems. The commonly used time correlation function is the KWW function, $\exp\left(-\left(\frac{t}{\tau}\right)^\alpha\right)$, which is a stretched exponential function. The time correlation function results from the random motion of particles, which can be modeled by rotational or translational diffusion. The time correlation function is a monoexponential function from normal diffusion, while it is a stretched exponential function or Mittag-Leffler function for anomalous diffusion, which could be described by the fractional rotational or translational diffusion proposed in Refs. [8,12]. The anomalous rotational diffusion can be modeled by [12, 30,31]

$$_tD_*^\alpha P = \frac{D_{f_r}}{a^\beta} \Delta_s^{\beta/2} P, \tag{45}$$

where $0 < \alpha, \beta \leq 2$, $D_{f_r}$ is the rotational diffusion coefficient, $a$ is the spherical radius, $_tD_*^\alpha$ is the Caputo fractional derivative defined by [19,20]

$$_tD_*^\alpha f(t) := \begin{cases} \frac{1}{\Gamma(m-\alpha)} \int_0^t \frac{f^{(m)}(\tau)d\tau}{(t-\tau)^{\alpha+1-m}}, & m-1 < \alpha < m, \\ \frac{d^m}{dt^m} f(t), & \alpha = m, \end{cases} \tag{46}$$

and $\Delta^{\beta/2}$ is the symmetric Riesz space-fractional derivative operator [19,20]. The time correlation funtion solved from Eq. (45) is [12]

$$G(t) = E_{\alpha,1}\left(-\frac{t^\alpha}{t_r}\right), \tag{47}$$

where

$$\tau_r = \left(\frac{1}{\frac{\beta}{6^2} \times \frac{D_{f_r}}{a^\beta}}\right)^{1/\alpha} \tag{48}$$

is the characteristic rotational time. The average phase diffusion jump time in the rotating frame based on Eq. (17) is

$$\tau_{jump}(\omega) = \int_{-\infty}^{\infty} E_{\alpha,1}(-|t|^\alpha/t_r) e^{-i\omega t} dt = \frac{2\omega^{\alpha-1}\tau_r^\alpha \sin(\pi\alpha/2)}{1+2(\omega\tau_r)^\alpha \cos(\pi\alpha/2)+(\omega\tau_r)^{2\alpha}}. \tag{49}$$

When $\alpha = 1$, $\beta = 2$, Eq. (47) reduces to Eq. (17), the normal rotational diffusion result.

In some systems, the time correlation function depends on the translation diffusion [1]. Anomalous diffusion can be modeled by [8,12]

$$_tD_*^\alpha P = 2D_{f_t}\Delta^{\beta/2}, \tag{50}$$

where $2D_{f_t}$ is the relative fractional translational diffusion coefficient of the two interactive spins, which have the same translational diffusion coefficient $D_{f_t}$. The correlation time is [12]

$$G^{(i)}(t) = \varepsilon^{(i)} \frac{N}{d^3} \int_0^\infty \left[J_{\frac{3}{2}}(u)\right]^2 E_{\alpha,1}\left(-\frac{2D_{ft}u^\beta t^\alpha}{d^\beta}\right) \frac{du}{u}, \tag{51}$$

where

$$\varepsilon^{(0)} = \frac{48\pi}{15}, \varepsilon^{(1)} = \frac{8\pi}{15}, \varepsilon^{(2)} = \frac{32\pi}{15},$$

$d$ is the distance, $u = kd$, $N$ is the density of spins, and $J_{\frac{3}{2}}$ is Bessel function [1,12]. The average phase jump time for translational diffusion based on Eq. (17) is

$$\begin{aligned}\tau_{jump}(\omega) &= 2\int_0^\infty G^{(i)}(t)e^{i\omega t}dt \\ &= \frac{\varepsilon^{(i)} N \tau_t^{1-\alpha}}{D_{ft}d^{3-\beta}(\omega\tau_t)^{1-\alpha}} \int_0^\infty \frac{\sin(\pi\alpha/2)u^{\beta-1}}{u^{2\beta}+2u^\beta(\omega\tau_t)^\alpha\cos(\pi\alpha/2)+(\omega\tau_t)^{2\alpha}}\left[J_{\frac{3}{2}}(u)\right]^2 du.\end{aligned} \tag{52}$$

These two $\tau_{jump}(\omega)$ in Eqs. (49) and (52) agree with the results in Ref. [12], and they can be used to obtain the effective diffusion constants in Eq. (18). The NMR relaxation time for dipolar coupling can then be obtained according to Eqs. (37-39), see Ref. [12] for a more detailed description of the fractional rotational and translational motion.

2.2 NMR relaxation described by anomalous phase diffusion

In the above, the diffusion is a normal diffusion where $\tau_{jump}(\omega)$ is assumed to be a constant, even with a fractional time correlation function, majorly because the Fourier transform in the rotating frame produces a finite average jump time. The normal phase diffusion proposed above could be an appropriate description of the phase evolution of spin systems. In a heterogeneous system, $\tau_{jump}(\omega)$ may not be a constant, and it could belong to a distribution behaving asymptotically to a power law. Additionally, $\gamma\hbar h_q$, the amplitude of the random field $H_q(t) = \gamma\hbar h_q G_q(t)$, could belong to a power-law distribution, which leads the phase jump length $\Delta\phi$ to follow a power-law distribution. Therefore, the phase diffusion could be anomalous [17,18], whose diffusion constant $D_{f\phi,lab}$ can be described by [16,19,20]

$$D_{f\phi,eff} \propto \frac{\langle|\Delta\phi|^\beta\rangle}{\langle\tau_{fjump}(\omega)\rangle^\alpha} \propto \frac{\omega_q^\beta \langle\tau_{fjump}(\omega)\rangle^\beta}{\langle\tau_{fjump}(\omega)\rangle^\alpha}, \tag{53}$$

where $\omega_q^\beta$ is proportional to $\langle\left|\frac{H_q(t)}{\hbar}\right|^\beta\rangle$, and $\tau_{fjump}(\omega)$ is the effective fractional mean jump time constant. The fractional phase diffusion equation can be described by [16,19,20]

$$_tD_*^\alpha P = D_{f\phi,eff}\Delta^{\beta/2}P, \tag{54}$$

where $0 < \alpha, \beta \leq 2$, $D_{f\phi,eff}$ is the fractional diffusion coefficient with units of rad$^\beta$/s$^\alpha$, $_tD_*^\alpha$ is the Caputo fractional derivative defined by Eq. (46), and $\Delta^{\beta/2}$ is the symmetric Riesz space-fractional derivative operator [19,20]. Performing Fourier transform on both sides of Eq. (54) gives

$$_tD_*^\alpha p(k,t) = -D_{f\phi,eff}\, k^\beta p(k,t), \tag{55}$$

and the signal amplitude $p(1,t)$ obeys

$$_tD_*^\alpha p(1,t) = -D_{f\phi,eff}\,p(1,t). \tag{56}$$

The evolution of the angular momentum $I_x$ and $I_z$, based on Eqs. (54), and (55), can be described as:

$$_tD_*^\alpha p_{Mx}(1,t) = -D_{f\phi2,eff}\,p_{Mx}(1,t), \tag{57a}$$

$$_tD_*^\alpha p_{Mz}(1,t) = -D_{f\phi1,eff}\bigl(p_{Mz}(1,t) - p_{Mz0}(1,0)\bigr). \tag{57b}$$

In Eq. (57b), the $p_{Mz0}(1,0)$ is included to account for the equilibrium magnetization of the thermal relaxation as mentioned in the obtaining of Eq. (25). Eq. (57b) implies that the random phase jump in every instant is modified by the thermal relaxation from the lattice. Replacing $P_{Mx}(1,t)$ and $P_{Mz}(1,t)$ with the magnetization $M_x(t)$ and $M_z(t)$ in Eqs. (57a) and (57b) respectively, we have:

$$_tD_*^\alpha M_x(t) = -\frac{1}{T_{2x}} M_x(t), \tag{58a}$$

where

$$T_{2x} = \frac{1}{D_{f\phi 2,eff}}; \tag{58b}$$

and

$$_tD_*^\alpha M_z(t) = -\frac{1}{T_1}(M_z(t) - M_{z0}), \tag{59a}$$

where

$$T_1 = \frac{1}{D_{f\phi 1,eff}}. \tag{59b}$$

Eqs. (58) and (59) are anomalous relaxation equations. These two equations agree with the results in Ref. [15]. The solutions of Eqs. (58) and (59) are:

$$M_x(t) = M_x(0) E_{\alpha,1}\left(-\frac{t^\alpha}{T_2}\right), \tag{60a}$$

$$M_z(t) = M_{z0} + (M_z(0) - M_{z0}) E_{\alpha,1}\left(-\frac{t^\alpha}{T_1}\right). \tag{60b}$$

## 3. Results and discussion

A phase diffusion equation method is proposed to explain the NMR spin-lattice and spin-spin relaxation processes. This method treats the spin system's evolution affected by the random field as a random walk process, which can be described by the normal or fractional phase diffusion equation. The method provides an intuitive physical explanation of the NMR relaxation, particularly for the complicated fractional NMR relaxation [15], which is challenging to be described by conventional perturbation theory [1,2].

Both the relaxation results for simple random field and dipolar coupling agrees with reported results [1,2]. From solving the normal phase diffusion equation, it is found that the relaxation rate is $W = \frac{1}{T_1} = D_\phi \propto \left(\frac{H_x(t)}{\hbar}\right)^2$, agreeing with the result of the time-dependent perturbation theory [1,2]. The obtained relaxation Eqs. (28) and (30) from the simple random perturbing field $H_1(t) = \sum_{q=x,y,z} H_q(t) I_q$ is the same as that obtained based on the Redfield theory presented in Ref. [2]. Additionally, the relaxation expression Eq. (39) for dipolar coupling replicates the results obtained by the operator method [1,2]. Although the results are consistent, the phase diffusion approach proposed here provides a more vivid physical picture of the evolution of the spin system compared to the complex quantum perturbation method.

The fractional phase diffusion can be model by different time derivatives. In the fractional diffusion equations (45), (50), and (54), the Caputo fractional derivative $_tD_*^\alpha$ is used, yielding Mittag-Leffler-function-based results. Another type of time-fractional derivative, the time fractal derivative $\frac{\partial}{\partial t^\alpha}$, has been applied to model various anomalous diffusion [32,33,34]. If $_tD_*^\alpha$ is replaced with $\frac{\partial}{\partial t^\alpha}$, the corresponding results will be changed from Mittag-Leffler function to stretched exponential function: $E_{\alpha,1}\left(-\frac{t^\alpha}{t_r}\right)$ in Eq. (47) and $E_{\alpha,1}\left(-\frac{2D_{ft}u^\beta t^\alpha}{d^\beta}\right)$ in Eq. (51) will be replaced with $\exp\left(-\frac{t^\alpha}{t_r}\right)$ and $\exp\left(-\frac{2D_{ft}u^\beta t^\alpha}{d^\beta}\right)$ respectively, and $E_{\alpha,1}\left(-\frac{t^\alpha}{T_2}\right)$ and $E_{\alpha,1}\left(-\frac{t^\alpha}{T_1}\right)$ in Eqs. (60a) and (60b) will be replaced with $\exp\left(-\frac{t^\alpha}{T_2}\right)$ and $\exp\left(-\frac{t^\alpha}{T_1}\right)$, respectively. When $x$ is small, the Mittag-Leffler function $E_{\alpha,1}(-x^\alpha)$ can be approximated as a stretched exponential function $\exp(-x^\alpha/\Gamma(1+\alpha))$. The frequently used time correlation function KWW function in polymer systems is a stretched exponential function.

Two diffusion processes affect NMR relaxation: the spatial rotational or translational diffusion due to the relative motion of lattice particles, and the phase diffusion of magnetization. The spatial diffusion modifies the Hamiltonian interaction exerting a random field to cause the phase diffusion, and it determines the time correlation function. For non-zero order quantum operator of like spins, the $\tau_{jump}(\omega)$ is a finite value in the rotating frame because the continuously rotating of spin moments at Larmor frequency makes the contribution of the Hamiltonians from its long time part average to zero.

However, the dependence of $\tau_{jump}(\omega)$ on $\tau$ has a significant distinction between normal spatial diffusion and anomalous spatial diffusion, which leads to different spin-lattice relaxation $T_1$ dependence on $\tau$. Figures 1 demonstrate the difference between the $^{13}$C $T_1$ values predicated according to Eq. (44c) from two types of $\tau_{jump}(\omega)$ based on MLF Eq. (49) and monoexponential function Eq. (17). The parameters used in the prediction include $n_H \frac{1}{15r^6}(\frac{\mu_0}{4\pi}\gamma_I\gamma_S)^2\hbar^2 S(S+1) = 2.28 \times 10^9$, and the two frequencies: 50.3 MHz and 100.6 MHz for $^{13}$C NMR. There are significant distinctions between the MLF and the monoexponential predicted curves. For fast motion with small $\tau$, the $T_1$ values based on mono exponential function are independent from magnetic field strength, which is often observed in a liquid NMR sample. In contrast, the $T_1$ values based on MLF strongly depend on the magnetic field strength. Additionally, unlike the sharp turn appearing in monoexponential curves, the MLF has smoothly bending curves. Such a field intensity dependence and the smooth bending of the curves are often observed in polymer samples.

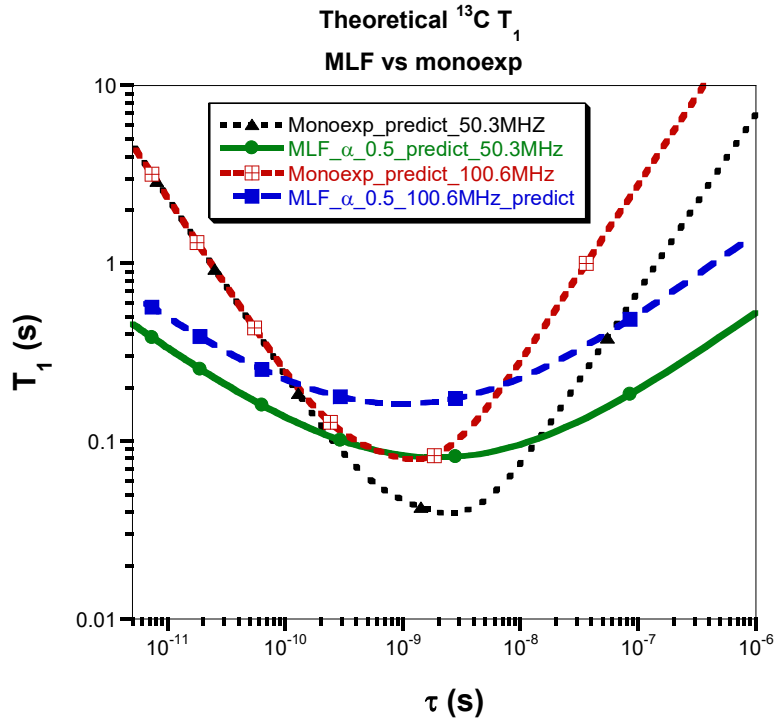

**Fig. 1** The difference between the $T_1$ values of $^{13}$C NMR predicated according to Eq. (44c) from two types of $\tau_{jump}(\omega)$ based on MLF Eq. (49) and monoexponential function Eq. (17). The parameters used in the prediction are $n_H \frac{1}{15r^6}(\frac{\mu_0}{4\pi}\gamma_I\gamma_S)^2\hbar^2 S(S+1) = 2.28 \times 10^9$, $n_H = 2$. The two field frequencies are 50.3 MHz and 100.6 MHz.

In Figure 2, the Eq. (44b) with $\tau_{jump}(\omega)$ based on MLF Eq. (49) is used to fit the experimental $^{13}$C $T_1$ data taken from Ref. [6], observed from methylene group of polyisobutylene (PIB) in 70% PIB and 30% head-to-head poly(propylene) (hhPP) from variable temperatures and two field frequencies, 50.3 MHz and 100.6 MHz for $^{13}$C NMR. The data cannot be interpreted by the monoexponential time correlation function. KWW function is often applied to fit such experimental data [6]. The $T_1$ based on MLF (Eqs. (44) and (49)) can successfully fit experimental $^{13}$C data, as demonstrated in Figure 2. In the fitting, the Vogel-Tamman-Fulcher (VTF) temperature dependence: $\tau_r = \tau_\infty \times 10^{\frac{B}{T-T_0}}$ is used to give the temperature-dependent segmental dynamics. The fitting parameter is listed in Table 1. The MLF based fitting gives similar results to that of the modified KWW (mKWW) function fitting reported in Ref. [6]. However, only four parameters are used in MLF fitting. In contrast, mKWW fitting needs six parameters, $\alpha$, $\tau_\infty$, $B$, $T_0$, $a_{lib}$, and $\tau_{lib}$. In Ref. [12], MLF based $^2$H $T_1$ has also been applied to fit $^2$H spin-lattice relaxation data of deuterium-labeled PIB/hhPP blend. No liberational motion is needed in both the MLF fitting presented in this paper and that reported in Ref. [12]. This may be because the Mittag-Leffler type time correlation function is equivalent to the combination of the slow segmental motion and the fast liberational motion [12]. MLF based $T_1$ can potentially be a suitable way to interpret the complex dynamics in polymer or biological systems.

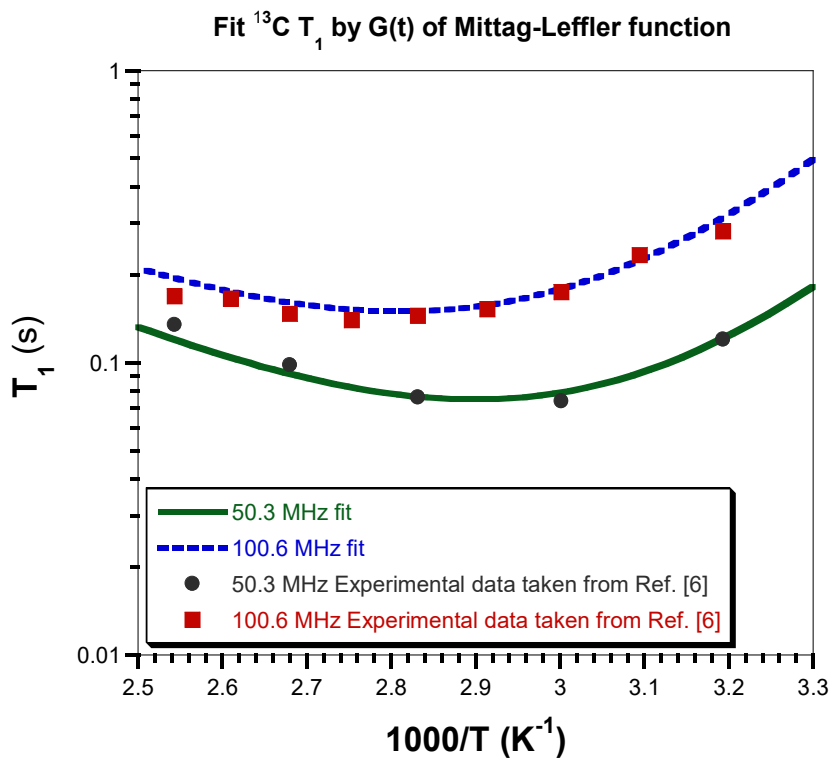

**Fig. 2** Fit the NMR experimental $^{13}$C $T_1$ data by Eq. (44b) with $\tau_{jump}(\omega)$ based on MLF Eq. (49). The experimental data are taken from Ref. [6], observed from methylene group of polyisobutylene (PIB) in 70% PIB and 30% head-to-head poly(propylene) (hhPP) from variable temperatures and two different field frequencies, 50.3 MHz and 100.6 MHz for $^{13}$C NMR.

**Table 1:** $^{13}$C Fitting parameters.

| Dynamic Mode | A | $\tau_\infty$ (ps) | B (K) | $T_0$ (K) |
|---|---|---|---|---|
| Eqs. (44) and (49) | 0.54 | 0.0046 | 1075 | 154 |
| mKWW parameters taken from Ref. [6] $\tau_{lib} = 0.1$ ps, $a_{lib} = 0.26$ | 0.6 | 0.1 | 775 | 160 |

While the kind of phase diffusion determines whether the relaxation equations are normal (Eqs.(26) and (30)) or anomalous (Eqs. (58-60)). The normal relaxation function is based on monoexponential function, while the anomalous relaxation function is either a Mittag-Leffler based function or a stretched exponential based function if the time fractal derivative $\frac{\partial}{\partial t^\alpha}$ is used. The difference between different types of relaxations is demonstrated by theoretical curves in Figures 3a and 3b. The parameters used in Figure 3a and 3b are $T_1 = 1$ s and $T_2 = 1$ s, respectively. The curves in Figure 3a shows that the longitudinal magnetization based on MLF *increases* relatively faster at a shorter time but slower at a longer time than the SEF based $T_1$ and monoexponential based $T_1$. While the curves in Figure 3b shows that the transverse magnetization based on MLF *decreases* relatively faster at a shorter time and slower at a longer time. Figure 3 indicates that the relaxation results from the same temperature could determine which types of NMR relaxation is present. In comparison, the anomalous rotational or translational diffusion determines the correlation time. To determine which type of rotational or translational diffusion is present, relaxation results from various temperatures and multiple magnetic fields are often necessary, as demonstrated in Figures 1 and 2.

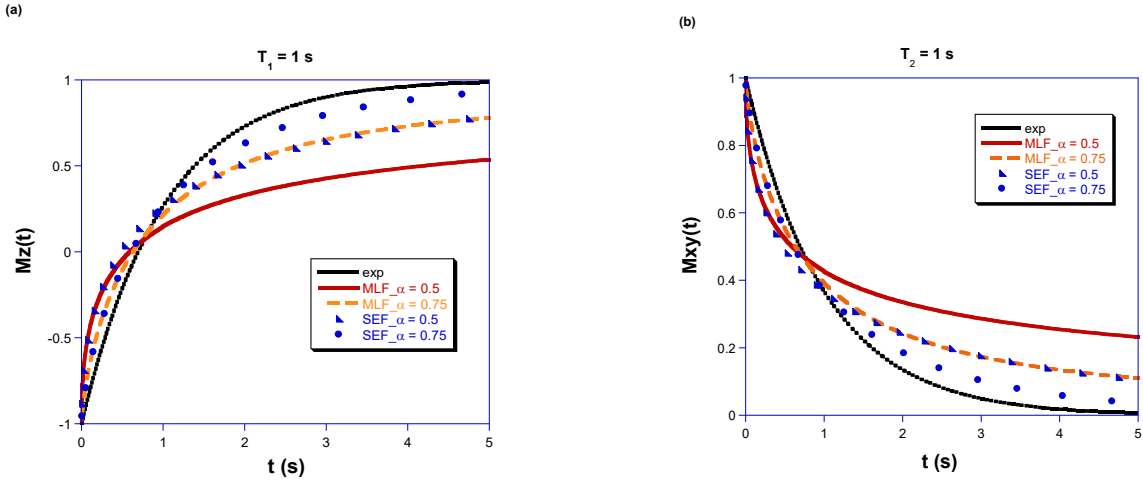

**Fig. 3** Theoretical difference between the relaxations of monoexponential function (exp), Mittag-Leffler function (MLF), and stretched exponential function (SEF): (a) spin-lattice relaxation, (b) spin-spin relaxation. The MLF relaxation is based on Eqs.(60a) and (60b)) for $T_1$ and $T_2$, respectively. Replacing the MLF in Eqs.(60a) and (60b) with monoexponential function and SEF function gives the monoexponential and SEF relaxation expressions. The parameters used are $T_1 = 1$ s and $T_2 = 1$ s, respectively.

The relaxation rate could yield an interesting traditional result. From Eq. (17), a simple monoexponential correlation function results in $\tau_{jump}(\omega) = \frac{2\tau}{1+(\omega\tau)^2}$ which implies $\tau_{jump}(\omega) = 1/\omega$ when the perturbation is on resonance. Therefore, the on resonance relaxation rate is $W = D_\phi = \frac{\omega_q^2 \tau_{jump}(\omega)}{2} = \frac{\omega_q^2}{2\omega} = \frac{H_q^2}{2\hbar} \cdot \frac{1}{\omega}$, and thus $\frac{W_1}{W_2} = \left(\frac{H_{1q}(t)}{H_{2q}(t)}\right)^2 \frac{\hbar\omega_2}{\hbar\omega_1}$. If this equation is possible to describe the relaxation of the combination state of spin and lattice relaxation where $\hbar\omega = E_S + E_L$, and the perturbation energies $H_{1q}(t)$ and $H_{2q}(t)$ are assumed to be the same, one has $\frac{W_1}{W_2} = \frac{\hbar\omega_2}{\hbar\omega_1} = 1 - \frac{\hbar\omega_1 - \hbar\omega_2}{\hbar\omega_1} = 1 - \frac{E_{S1} - E_{S2}}{E_{S1} + E_L}$. When $E_{S1} - E_{S2}$ is small compared to $E_{S1} + E_L$, $\frac{W_1}{W_2} \cong \exp(-\frac{E_{S1} - E_{S2}}{E_{S1} + E_L})$. Because the energy $E_{S1} + E_L$ could be replaced with $KT$, we could have $\frac{W_1}{W_2} = \exp(-\frac{E_{S1} - E_{S2}}{KT})$, which leads to a Boltzmann distribution of the system [1,2]. Note both $W$ and $\omega$ mentioned in this paragraph are for the whole spin and lattice system, rather than spin only.

In general, the phase diffusion approach is based on a classical method (the evolution of the magnetization, phase diffusion), although certain quantum mechanics concepts such as the angular momentum operator and spin coherences are employed. The application of a classical method to describe the spin system is reasonable. Ref. [2] shows that the expectation values of spin moments derived from quantum mechanics obey a classical equation, $\frac{d\langle\mu\rangle}{dt} = \langle\mu\rangle \times \gamma H$, which works for a single spin, or a bulk magnetization (when the investigated spins are statistically independent). The established traditional quantum density operator method is more convenient to calculate the complex Hamiltonian interaction than the phase diffusion method. However, the diffusion method proposed here gives a clearer visual, and it provides a convenient way to understand the anomalous relaxation process that is challenging for the established methods. Additionally, various related dynamic parameters can be analyzed based on the phase diffusion method, such as the anomalous relaxation time, $\frac{1}{T_1} = D_{f\phi,eff} \propto \frac{\langle|\Delta\phi|^\beta\rangle}{\langle\tau_{fjump}(\omega)\rangle^\alpha}$ from this method. Further effort is needed to improve the current method and to better understand the anomalous NMR relaxation and other complicated Hamiltonians such as quadrupolar interaction and anisotropic chemical shift.

**Appendix A  Effect of Thermal Equilibrium of Lattice**

The spin system is a part of the large system including both the spin and the lattice. Regardless of the spin relaxation, one can assume that the lattice is still in thermal equilibrium. When the associated lattice transition is considered, the transition rates $W_{\alpha,\beta}$ and $W_{\beta,\alpha}$ are different, and [1,2]

$$\frac{W_{\beta,\alpha}}{W_{\alpha,\beta}} = \exp\left(\frac{\hbar\omega}{KT}\right), \qquad (A.1)$$

which can be explained by the factor $\langle|\langle F^{(2)}(\Omega)\rangle|^2\rangle$ is different by $\exp\left(\frac{\hbar\omega}{KT}\right)$, when $\langle|\langle F^{(2)}(\Omega)\rangle|^2\rangle$ is average over two different sets of the lattice states, $f, f'$ and $f', f$ [1]. Eq. (A.1) can also be obtained based on the transition rate at the equilibrium state [2]. For a simple two-level system with states $\alpha$ and $\beta$, the transition rate equations are [2]

$$\frac{dp_{z\alpha}(1,t)}{dt} = -D_{\phi x,eff,\alpha\beta} p_{z\alpha}(1,t) + D_{\phi x,eff,\beta\alpha} p_{z\beta}(1,t),$$
$$\frac{dp_{z\beta}(1,t)}{dt} = -D_{\phi x,eff,\beta\alpha} p_{z\beta}(1,t) + D_{\phi x,eff,\alpha\beta} p_{z\alpha}(1,t), \qquad (A.2)$$

where $p_{z\alpha}(1,t) = \langle I_z\rangle_\alpha$, $p_{z\beta}(1,t) = \langle I_z\rangle_\beta$, $D_{\phi x,eff,\alpha\beta}$, and $D_{\phi x,eff,\beta\alpha}$ corresponds to the effect of perturbations $I^+$ and $I^-$ respectively. Eq. (A.2) leads to [2]

$$\frac{dp_z(1,t)}{dt} = -D_{\phi x,eff}(p_z(1,t) - p_{z0}(1,0)), \qquad (A.3)$$

where

$$p_z(1,t) = p_{z\beta}(1,t) - p_{z\alpha}(1,t),$$
$$D_{\phi x,eff} = D_{\phi x,eff,\alpha\beta} + D_{\phi x,eff,\beta\alpha},$$

$$p_{z0}(1,0) = \left(p_{z\alpha}(1,t) + p_{z\beta}(1,t)\right)\frac{D_{\phi x,eff,\beta\alpha}-D_{\phi x,eff,\alpha\beta}}{D_{\phi x,eff,\alpha\beta}+D_{\phi x,eff,\beta\alpha}}. \tag{A.4}$$

Because of the high-temperature approximation ($\exp\left(\frac{\hbar\omega}{KT}\right) \cong 1$) [1,2], $D_{\phi x,eff,\alpha\beta}$ and $D_{\phi x,eff,\beta\alpha}$ are approximately half of $D_{\phi x,eff}$ used in this paper. This approximation is reasonable because when the two conjugate operators such as $I^+$, $I^-$ are considered individually, their corresponding rate is half of that considered together. In the other part of the paper, the rate consider the two conjugate terms together, and $\langle I_z \rangle = \langle I_z \rangle_\beta - \langle I_z \rangle_\alpha$. Therefore, the results for the relaxation are not affected. In Ref. [1], it is possible to derive the thermal equilibrium correction in another way, which needs further research.

Similar to the above process, it is easy to obtain

$$_tD_*^\alpha p_z(1,t) = -D_{f\phi,eff}\left(p_z(1,t) - p_{z0}(1,0)\right), \tag{A.5}$$

where

$$p_z(1,t) = p_{z\beta}(1,t) - p_{z\alpha}(1,t),$$
$$D_{f\phi,eff}=D_{f\phi,eff,\alpha\beta} + D_{f\phi,eff,\beta\alpha},$$
$$p_{z0}(1,0) = \left(p_{z\alpha}(1,t) + p_{z\beta}(1,t)\right)\frac{D_{f\phi,eff,\beta\alpha}-D_{f\phi,eff,\alpha\beta}}{D_{f\phi,eff,\alpha\beta}+D_{f\phi,eff,\beta\alpha}}. \tag{A.6}$$